\documentclass{svmult}

\begin{document}

\title*{The Strong CP Problem and Axions}

\author{R.D. Peccei}

\institute{Department of Physics and Astronomy, UCLA, Los Angeles, California, 90095
\texttt{peccei@physics.ucla.edu}}

\maketitle

\begin{abstract}
I describe how the QCD vacuum structure, necessary to resolve the
$U(1)_A$ problem, predicts the presence of a P, T and CP violating term
proportional to the vacuum angle $\bar{\theta}$. To agree with
experimental bounds, however, this parameter must be very small
$(\bar{\theta} \leq 10^{-9}$). After briefly discussing some possible
other solutions to this, so-called, strong CP problem, I concentrate on
the chiral solution proposed by Peccei and Quinn which has associated
with it a light pseudoscalar particle, the axion. I discuss in detail
the properties and dynamics of axions, focusing particularly on
invisible axion models where axions are very light, very weakly coupled
and very long-lived. Astrophysical and cosmological bounds on invisible
axions are also briefly touched upon.
\end{abstract}

\section{ The $U(1)_A$ Problem and Its Resolution}

In the 1970's the strong interactions had a puzzling problem, which became particularly clear with the development of QCD. The QCD Lagrangian for N flavors in the limit of vanishing quark masses $ m_f \to 0$ has a large global symmetry: $U (N)_V \times U (N)_A$. Since $m_u, m_d << \Lambda_{QCD} $, we know that, at least for these quarks, the limit of sending the quark masses to zero is sensible. Thus one would expect the strong interactions to be approximately $U (2)_V \times U (2)_A$ invariant.

What one finds experimentally is that, indeed, the vector symmetry corresponding to isospin times baryon number [$U(2)_V = SU(2)_I \times U(1)_B$] is a good approximate symmetry of nature, as manifested by the appearance of nucleon and pion multiplets in  the spectrum of hadrons. For axial symmetries, however, things are different. Dynamically, quark condensates $ <\bar{u}u>=<\bar{d}d> \neq 0$ form, breaking the axial symmetry down spontaneously. So one does not expect approximate mixed parity multiplets in the hadronic spectrum, but vestiges of the four Nambu-Goldstone bosons associated with the breakdown of $U(2)_A$. Although pions are light, $m_{\pi} \simeq 0 $, there are no signs of  another light state in the hadronic spectrum, since  $m^2_{\eta}>> m^2_{\pi}$. Weinberg \cite{Weinberg} dubbed this the $U(1)_A$ problem and suggested that, somehow, there was no $U(1)_A$ symmetry in the strong interactions.

The resolution of the $U(1)_A$ problem came through the realization by 't Hooft \cite{'t Hooft} that the QCD vacuum has a more complicated structure. The more complex nature of the QCD vacuum, in effect, makes $U(1)_A$ not a true symmetry of QCD, even though it is an apparent symmetry of the QCD Lagrangian in the limit of vanishing quark masses. However, associated with this more complicated QCD vacuum there is a phase parameter $\theta$ and only if this parameter is very small is CP not very badly broken in the strong interactions. So the solution of the $U(1)_A$ problem begets a different problem: why is CP not badly broken in QCD? This is known as the strong CP problem.

A possible resolution of the $U(1)_A$ problem seems to be provided by the chiral anomaly for axial currents. \cite{ABJ}  The divergence of the axial current $J^{\mu}_5$ associated with $U(1)_A$ gets quantum corrections from the triangle graph which connects it to two gluon fields with quarks going around the loop. This anomaly gives a non-zero divergence for $J^{\mu}_5$
\begin{equation}
\partial_{\mu} J^{\mu}_5= \frac{g^2N}{32\pi^2} F^{\mu\nu}_a\tilde{F}_{a\mu\nu},
\end{equation}
where $\tilde{F}_{a\mu\nu}= \frac{1}{2}\epsilon_{\mu\nu\alpha\beta}F^{\alpha\beta}_a$. Hence, in the massless quark limit, although formally QCD is invariant under a $U(1)_A$ transformation 
\begin{equation}
q_f \to e^{i\alpha \gamma_5/2}q_f
\end{equation}
the chiral anomaly affects the action:
\begin{equation}
\delta W= \alpha \int d^4x \partial_{\mu} J^{\mu}_5= \alpha\frac{g^2N}{32\pi^2}\int d^4x F^{\mu\nu}_a\tilde{F}_{a\mu\nu}.
\end{equation}

However, matters are not that simple! The pseudoscalar density entering in the anomaly is, in fact, a total divergence \cite{Bardeen}: 
\begin{equation}
F^{\mu\nu}_a\tilde{F}_{a\mu\nu}= \partial_{\mu} K^{\mu},
\end{equation}
where
\begin{equation}
K^{\mu}= \epsilon^{\mu \alpha \beta\gamma}A_{a\alpha}[F_{a\beta \gamma} -\frac{g}{3} f_{abc} A_{b\beta} A_{c\gamma}].
\end{equation}
Because of these identities $\delta W$ is a pure surface integral
\begin{equation}
\delta W= \alpha\frac{g^2N}{32\pi^2}\int d^4x \partial_{\mu}K^{\mu}= \alpha\frac{g^2N}{32\pi^2}\int d\sigma_{\mu} K^{\mu}.
\end{equation}
Hence, using the naive boundary condition  that $A_a^{\mu}=0$ at spatial infinity, one has $\int d\sigma_{\mu} K^{\mu}=0$  and $U(1)_A$ appears to be a symmetry again. 
What 't Hooft \cite{'t Hooft} showed, however, is that the correct boundary condition to use is that $A_a^{\mu}$  should be a pure gauge field at spatial infinity (i.e. either $A_a^{\mu} =0$, or a gauge transformation of 0). It turns out that, with these boundary conditions there are gauge configurations for which $\int d\sigma_{\mu} K^{\mu}\neq 0$ and thus $U(1)_A$ is not a symmetry of QCD.

This is most easily understood by working in the $A^o_a$ gauge. Studying $SU(2)$ QCD for simplicity, in this gauge \cite{CDG} one has only spatial gauge fields $A^i_a$. Under a gauge transformation these fields transform as:
\begin{equation}
\frac{1}{2}\tau_aA^i_a \equiv A^i \to \Omega A^i \Omega^{-1} + \frac{i}{g} \nabla^i \Omega \Omega^{-1}.
\end{equation}
Thus vacuum configurations either vanish or have the form $\frac{i}{g} \nabla^i \Omega \Omega^{-1}$ In the $A^o_a=0$ gauge one can further classify these vacuum configurations by how $\Omega$ goes to unity as $\vec{r} \to \infty$:
\begin{equation} 
 \Omega_n \to e ^{i2\pi n}  ~~ \rm{as}~~ \vec{r} \to \infty~~  [n=0, \pm1, \pm 2,...].
\end{equation}
The integer $n$ (the winding number) is related to the Jacobian of an $S_3 \to S_3$ map and is given by \cite{Crewther}
\begin{equation}
n=\frac{ig^3}{24\pi^2} \int d^3r Tr~ \epsilon_{ijk} A^i_nA^j_nA^k_n.
\end{equation}
This expression is closely related to the Bardeen current $K^{\mu}$. Indeed, in the $A^o_a=0$ gauge  only $K^o\neq 0$ and one finds for pure gauge fields:
\begin{equation}
     K^o=-\frac{g}{3}\epsilon_{ijk}\epsilon_{abc} A^i_a A^j_b A^k_c =\frac{4}{3}ig \epsilon_{ijk}Tr ~A^i A^j A^k.
\end{equation}

The true vacuum is a superposition of these, so-called, n-vacua and is called the $\theta$-vacuum: 
 \begin{equation}                  
 |\theta> = \Sigma_n e^{ -in\theta} |n>.
\end{equation}
It is easy to see that in the vacuum to vacuum transition amplitude there are transitions with $\int d\sigma_{\mu}K^{\mu} \neq 0$. Indeed,
\begin{equation}
  n|_{t= +\infty} - n|_{t= -\infty}  =  \frac{g^2}{32\pi^2 } \int d \sigma_{\mu}K^{\mu}|^{t=+\infty}_{ t= -\infty}.
\end{equation}

Using Eq (12), in detail one can write for the vacuum to vacuum transition amplitude
\begin{equation}
_+<\theta|\theta>_- = \Sigma_{m,n} e^{im\theta} e^{ -in\theta}~ _+<m|n>_- = \Sigma_{\nu}e^{i\nu\theta} \Sigma_n ~ _+<n+\nu|n>_-.
\end{equation}
It is easy to see that the difference in winding numbers $\nu$ is given by
\begin{equation}
\nu=\frac{g^2}{32\pi^2 } \int d \sigma_{\mu}K^{\mu}|^{t=+\infty}_{ t= -\infty} =\frac{g^2}{32\pi^2 } \int d^4x F^{\mu\nu}_a\tilde{F}_{a\mu\nu}
\end{equation}
Using the usual path integral representation for the vacuum to vacuum amplitude $_+<\theta|\theta>_-$, one sees that
\begin{equation}
_+<\theta|\theta>_- =\Sigma_{\nu}\int \delta Ae^{iS_{\rm{eff}}[A]} \delta[\nu- \frac{g^2}{32\pi^2 } \int d^4x F^{\mu\nu}_a\tilde{F}_{a\mu\nu}],
\end{equation}
where
\begin{equation}
S_{\rm{eff}}[A] = S_{\rm{QCD}}[A] + \theta \frac{g^2}{32\pi^2 } \int d^4x F^{\mu\nu}_a\tilde{F}_{a\mu\nu}.
\end{equation}

The resolution of $U(1)_A$ problem, by recognizing the complicated nature of the QCD's vacuum, effectively adds and extra term to the QCD Lagrangian:
\begin{equation}
L_{\theta} =\theta \frac{g^2}{32\pi^2 }  F^{\mu\nu}_a\tilde{F}_{a\mu\nu}.
\end{equation}
This term violates Parity and Time reversal invariance, but conserves Charge conjugation invariance, so it violates CP. The existing strong bound on the neutron electric dipole moment \cite{bound} $|d_n|< 3 \times 10^{-26}$ ecm requires the angle $\theta$ to be very small [$d_n\simeq e \theta m_q/M_N^2 $ implies \cite{Baluni,CDVW} $\theta < 10^{-9}$]. Why should this be so is known as the {\bf{strong CP problem}}.

This problem is actually worse if one considers the effect of chiral transformations on the $\theta$-vacuum. Chiral transformations, because of the anomaly, actually can change the $\theta$-vacuum \cite{JR}:
\begin{equation}
 e^{i\alpha Q_5} | \theta > = | \theta + \alpha> .
\end{equation}
If, besides QCD, one includes the weak interactions, the quark mass matrix is in general complex:
\begin{equation}
L_{\rm{Mass}} = \bar{q}_{iR} M_{ij} q_{jL} + h. c.
\end{equation}
To go to a physical basis one must diagonalize this mass matrix and when one does so, in general, one performs a chiral transformation which changes $\theta$ by $ Arg~ Det M$. So, in the total theory, the coefficient of the $ F\tilde{F}$ term is

\begin{equation}
\bar{\theta}= \theta + Arg ~det M
\end{equation}
The strong CP problem is why is this $\bar{\theta}$ angle, coming from the strong and weak interactions, so small?

\section{Approaches to the Strong CP Problem}

There are three possible "solutions" to the strong CP Problem:

  \noindent i.            Unconventional dynamics

    \noindent ii.            Spontaneously broken CP

 \noindent iii.            An additional chiral symmetry

\noindent However, in my opinion, only the third of these is a viable solution. Of course, it might be possible that, as a result of some anthropic reasons $\bar{\theta} $ just turns out to be of $O(10^{-10})$ but I doubt it, as a Universe where CP is violated strongly seems as viable as one where it is not. \cite{Wilczek}

Appealing to unconventional dynamics to solve the strong CP problem is also not very believable. Attempts have been made to suggest that the boundary conditions which give rise to the $\theta$-vacuum are an artifact \cite{arti}, but then what is the solution to the $U(1)_A $ problem? Other approaches try to use the periodicity of vacuum energy $E(\theta) \sim cos \theta$ to deduce that $\theta$ vanishes\cite{period}, but fail to motivate why one should minimize the vacuum energy.

The second possibility, of spontaneously broken CP, is more interesting. Obviously, if CP is a symmetry of nature which is spontaneously broken, then one can set $\theta=0$ at the Lagrangian level. However, if CP is spontaneously broken $\theta$ gets induced back at the loop-level. To get $\theta < 10^{-9}$ one needs, in general, to insure that $\theta$ vanishes also at the 1-loop level. Although models exist where this is accomplished \cite{models}, theories with spontaneously broken CP need complex Higgs VEVs, leading to difficulties with FCNC and domain walls \cite{KOZ} and  one must introduce recondite physics \cite{Barr, Nelson} to avoid these problems. In my view, however, the biggest drawback for this "solution" to the strong CP problem is that experimental data is in excellent agreement with  the CKM Model- a model where CP is explicitly, not spontaneously broken.

Introducing an additional chiral symmetry is a very natural solution for the strong CP problem since this chiral symmetry, effectively, rotates the $\theta$ -vacua away. Two suggestions have been made for this chiral symmetry:

 \noindent     i.            The  u-quark has no mass, $m_u = 0$ \cite{mu}

  \noindent   ii.            The Standard Model has an additional global U(1) chiral symmetry \cite{PQ}

\noindent The first possibility is disfavored by a standard current algebra analysis \cite{ Leutwyler} which shows that all data is consistent with a finite mass for the u-quark. Furthermore, it is difficult to understand what  would be the origin of this chiral symmetry, which asks effectively that $ Arg~ det M = 0$.

\section{$U(1)_{PQ}$ and Axions}

Introducing a global chiral $U(1)$ symmetry \cite{PQ}- which has become known as a $U(1)_{PQ}$ symmetry- provides perhaps the most cogent solution to the strong CP problem. This symmetry is necessarily spontaneously broken, and its introduction into the theory effectively replaces the static CP-violating angle $\bar{\theta}$ with a dynamical CP- conserving field- the axion. The axion is the Nambu- Goldstone boson of the broken $U(1)_{PQ}$ symmetry. \cite{WW} As a result, under a $U(1)_{PQ}$ transformation, the axion field $a(x)$ translates
\begin{equation}
a(x) \to a(x) + \alpha f_a,
\end{equation}
where $f_a$ is the order parameter associated with the breaking of $U(1)_{PQ}$.

Formally, to make the Lagrangian of the Standard Model $U(1)_{PQ}$ invariant this Lagrangian must be augmented by axion interactions: 
\begin{equation}
L_{\rm{total}}= L_{\rm{SM}} +\bar{\theta}\frac{g^2}{32\pi^2} F_a^{\mu \nu}\tilde{F}_{ a\mu \nu} -\frac{1}{2} \partial_{\mu}a\partial^{\mu}a 
+ L_{\rm{int}}[\partial^{\mu} a/f_a; \Psi] + \xi\frac{a}{f_a}\frac{g^2}{32\pi^2} F_a^{\mu \nu}\tilde{F}_{ a\mu \nu}. 
\end{equation}
The last term above is needed to ensure that the $U(1)_{PQ}$ current indeed has a chiral anomaly:
\begin{equation}
\partial_{\mu} J^{\mu}_{PQ}=\xi \frac{g^2}{32\pi^2} F_a^{\mu \nu}\tilde{F}_{ a\mu \nu}.
\end{equation}
This term also represent an effective potential for the axion field, and its minimum occurs at $<a>=-\frac{f_a}{\xi} \bar{\theta}$:
\begin{equation}
<\frac{\partial V_{\rm{eff}}}{\partial a}>= -\frac{\xi}{f_a}\frac{g^2}{32\pi^2}< F_a^{\mu \nu}\tilde{F}_{ a\mu \nu}>| _{<a>=-\frac{f_a}{\xi} \bar{\theta}}~~=0.
\end{equation}
Since at the minimun the $\bar{\theta}$-term is cancelled out, this provides a dynamical solution to the strong CP problem. \cite{PQ}
 
It is easy to understand the physics of the Peccei Quinn solution to the strong CP problem. If one neglects the effects of QCD then the extra $U(1)_{PQ}$ symmetry introduced allows all values for $<a>$ to exist: $0 \leq~ <a>~ \leq 2\pi$. However,
including the effects of the QCD anomaly serves to generates a potential for the axion field which is periodic in the effective vacuum angle $\bar{\theta} + \xi \frac{<a>}{f_a}$:
\begin{equation}
 V_{\rm{eff}}\sim cos[\bar{\theta} + \xi \frac{<a>}{f_a} ].
\end{equation}
Minimizing this potential with respect to $<a>$ gives the PQ solution:
\begin{equation}
<a>=-\frac{f_a}{\xi} \bar{\theta},
\end{equation}
Obviously, the Lagrangian (22) written in terms of $a_{\rm{phys}}= a- <a>$
no longer has a CP violating $\bar{\theta}$-term.

Expanding $V_{\rm{eff}}$ at the minimum gives the axion a mass
\begin{equation}
m_a^2= <\frac{\partial^2 V_{\rm{eff}}}{\partial a^2}>=-\frac{\xi}{f_a}\frac{g^2}{32\pi^2}\frac{\partial}{\partial a}< F_a^{\mu \nu}\tilde{F}_{ a\mu \nu}>| _{<a>=-\frac{f_a}{\xi} \bar{\theta}}
\end{equation}
The calculation of the axion mass was first done explicitly by current algebra techniques by Bardeen and Tye. \cite{BT} Here I will give an effective Lagrangian derivation \cite{BPY} of $m_a$, as this technique also gives readily the couplings of axions to matter.

\section{Axion Dynamics}

In the original Peccei Quinn model, the $U(1)_{PQ}$ symmetry breakdown coincided with that of electroweak breaking $f_a = v_F$, with $v_F \simeq $250 GeV. However, this is not necessary. If $f_a >> v_F$ then the axion is very light, very weakly coupled and very long lived. Models where this occurs have become known as invisible axion models.

It is useful to derive first the properties of weak-scale axions and then generalize the discussion to invisible axion models. To make the Standard Model (SM) invariant under a $ U(1)_{PQ }$ transformation one must introduce 2 Higgs fields to absorb independent chiral transformations of the  u- and d-quarks (and leptons). The relevant Yukawa interactions involving these Higgs fields in the SM are:
\begin{equation}
L_{\rm {Yukawa}}= \Gamma^u_{ij}\bar{Q}_{Li}\Phi_1u_{Rj} + \Gamma^d_{ij}\bar{Q}_{Li}\Phi_2d_{Rj}+ \Gamma^{\ell}_{ij}\bar{L}_{Li}\Phi_2\ell_{Rj} + h. c.
\end{equation}
Defining $x=v_2/v_1$ and $v_F= \sqrt{v_1^2 + v_2^2}$, the axion is the common phase field in $\Phi_1$ and $\Phi_2$ which is orthogonal to the weak hypercharge:
\begin{equation}
\Phi_1=\frac{v_1}{\sqrt{2}} e^{ia x/v _F}[\begin{array}{c}1\\0\\ \end{array}]~~;~~\Phi_2=\frac{v_2}{\sqrt{2}} e^{ia/ xv _F}[\begin{array}{c}0\\1\\ \end{array}].
\end{equation}
From the above, it is clear that $L_{\rm{Yukawa}}$ is invariant under the $U(1)_{PQ}$ transformation
\begin{equation} 
a \to a + \alpha v_F~; ~u_{Rj} \to e^{-i\alpha x} u_{Rj} ~;~ d_{Rj} \to e^{-i\alpha/x} d_{Rj}~;~ \ell_{Rj}\to e^{-i\alpha/x} \ell_{Rj}.
\end{equation}

Let us focus on the quark pieces. The symmetry current for $U(1)_{PQ}$,
\begin{equation}
J^{\mu}_{PQ}=-v_F \partial^{\mu}a + x \Sigma_i \bar{u}_{iR}\gamma^ {\mu} u_{iR} + \frac{1}{x} \Sigma_i \bar{d}_{iR}\gamma^ {\mu} d_{iR} 
\end{equation}
identifies the anomaly coefficient $\xi$ in Eq. (23) as:
\begin{equation}
\xi=\frac{N}{2}(x +1/x)=N_g(x +1/x).
\end{equation}
To compute the axion mass, it is useful to separate the effects of the axion interactions with the light quarks from the rest. These interactions can be deduced from the underlying theory by constructing an appropriate effective chiral Lagrangian. The effects of the heavy quarks essentially can  then be accounted through their contribution to the chiral anomaly of $ J^{\mu}_{PQ}$

For the two light quarks, as usual \cite{Leutwyler}, one introduces a $2\times 2$ matrix of Nambu-Goldstone fields 
\begin{equation}
\Sigma = exp[ i\frac{(\vec{\tau}.\vec{\pi} +\eta)}{f_{\pi}}],
\end{equation}
where $f_{\pi}$ is the pion decay constant.
Then the meson sector of the light quark theory, neglecting the effect of the Yukawa interactions, is embodied in the $U(2)_V\times U(2)_A$ invariant effective Lagrangian
\begin{equation}
L_{\rm {chiral}} =-\frac{f_{\pi}^2}{4} Tr \partial_{\mu}\Sigma \partial^{\mu}\Sigma^{\dagger}.
\end{equation}

To $ L_{\rm {chiral}} $ one must add $U(2)_V\times U(2)_A$ breaking terms which mimic the $U(1)_{PQ}$ invariant Yukawaš interactions of the u- and d-quarks. This is accomplished by the Lagrangian
\begin{equation} 
 L_{\rm{mass}}=\frac{1}{2}(f_{\pi} m^o_{\pi} )^2 Tr[\Sigma AM+ (\Sigma AM)^{\dagger}],
\end{equation}
where
\begin{equation}
A=[\begin{array}{cc}e^{-iax/v_F}&0\\0& e^{-ia/xv_F}\\ \end{array}]~~;~~
M=[\begin{array}{cc}\frac{m_u}{m_u+m_d}&0\\0& \frac{m_d}{m_u+m_d}\\ \end{array}].
\end{equation}
Note that the invariance of $ L_{\rm{mass}}$ under $U(1)_{PQ}$ requires that $\Sigma$ transform as
\begin{equation}
\Sigma \to \Sigma [\begin{array}{cc}e^{i\alpha x}&0\\0& e^{i\alpha/x}\\ \end{array}].
\end{equation}

$L_{\rm{mass}}$, however, only gives part of the physics associated with the symmetry breakdown of $ U(2)_A$. In fact, the quadratic terms in $ L_{\rm{mass}}$ involving neutral fields 
\begin{equation}
L^{(2)}_{\rm{mass}}=-\frac{ (m^o_{\pi})^2}{2}\{\frac{m_u}{m_u+m_d}[\pi^o+\eta-\frac{xf_{\pi}}{v_F}a]^2+ \frac{m_d }{m_u+m_d}[\eta -\pi^o-\frac{f_{\pi}}{xv_F}a]^2 \}
\end{equation}
give for the ratio
\begin{equation} 
\frac{m_{\eta}^2}{m_{\pi}^2}= \frac{m_d}{m_u} \simeq 1.6.
\end{equation}
which contradicts experiment. Indeed, if $L_{\rm{mass}}$ was all that there was, we would have recreated in the effective Lagrangian language the $U(1)_A$ problem! Furthermore, with only this term the axion is still massless.

The resolution of the $U(1)_A$ problem in the effective Lagrangian theory is achieved by adding a further mass term which takes into account of the anomaly in both $U(1)_A$ and $U(1)_{PQ}$. This mass term gives the $\eta$ the right mass and produces a mass for the axion. It is easy to convince oneself that such a term has the form \cite{BPY}
\begin{equation} 
 L_{\rm{anomaly}}=-\frac{(m^o_{\eta})^2}{2}  [\eta+ \frac{f_{\pi}}{v_F} \frac{(N_g-1)(x +1/x)}{2}a ]^2,
\end{equation}
where $( m^o_{\eta})^2 \simeq m_{\eta}^2>> m_{\pi}^2$. The
coefficient in front of the axion field in $L_{\rm{anomaly}}$ reflects the relative strength of the couplings of the axion and the $\eta$ što $ F\tilde{F}$ as the result of the anomalies in $U(1)_{PQ}$ and $U(1)_A$. Naively, the ratio of these couplings is just $f_{\pi}/2v_F \xi$. However, the reason that $N_g-1$ appears in the above, rather than $N_g$, is that $ L_{\rm{mass}}$ already includes the light quark interactions of axions, so only the contribution of heavy quarks to the PQ anomaly should be taken into account in $L_{\rm{anomaly}}$.

Diagonalization of the quadratic terms in $ L_{\rm{mass}}$ and $L_{\rm{anomaly}}$ gives both the axion mass and the parameters for axion-pion and axion-eta mixing for the PQ model. It is convenient to define
\begin{equation}
\bar{ m_a} = m_{\pi} \frac{f_{\pi}}{v_F}\frac{\sqrt{m_um_d}}{m_u+m_d} \simeq 25 ~{\rm keV}.
\end{equation}
Then one finds:
\begin{equation}
m_a= \lambda_m \bar{ m_a}~;~\xi_{a\pi}=\lambda_3 \frac{f_{\pi}}{v_F}~;\xi_{a\eta}=\lambda_0\frac{f_{\pi}}{v_F},
\end{equation}
where
\begin{eqnarray}
\lambda_m = N_g(x +1/x)\\ 
\lambda_3=\frac{1}{2}[(x -1/x)-N_g(x +1/x)\frac{m_d-m_u}{m_u+m_d}]\\
\lambda_0=\frac{1}{2}(1-N_g)(x+1/x).
\end{eqnarray}

In addition to the three parameters above, all axion models are characterizedš also by how the axion couples to two photons. Writing the interaction Lagrangian describing this coupling as
\begin{equation}
L_{a\gamma\gamma}= \frac{\alpha}{4\pi} K_{a\gamma\gamma} \frac{a_{\rm{phys.}}}{f_a} F^{\mu\nu} \tilde{F}_{\mu\nu}.
\end{equation}
one needs to find the coupling $K_{a\gamma\gamma}$for the PQ model. This coupling follows from the electromagnetic anomaly of the PQ current 
\begin{equation}
\partial_{\mu} J^{\mu}_{PQ}= \frac{\alpha}{4\pi}\xi_{\gamma}F_{\mu\nu}\tilde{F}^{\mu\nu}. 
\end{equation}
Here $\xi_{\gamma}$ gets contributions from both quarks and leptons, and one finds: 
\begin{equation} \xi_{\gamma}= N_g\{[3(2/3)^2]x+[3(-1/3)^2+(-1)^2]1/x\}
=\frac{4}{3}N_g(x +1/x). 
\end{equation}
As before, in computing $K_{a\gamma\gamma}$ one must separate out the light quark contribution of the axion in the anomaly ( so that $\xi_{\gamma}^{\rm{eff.}}= \frac{4}{3}N_g(x +1/x)-\frac{4}{3}x- \frac{1}{3}\frac{1}{x}$) and add back the axion to two-photon contribution which arises from the coupling of the $\pi^o$ and $\eta$ to two photons, via axion-pion and axion-eta mixing: $\lambda_3 +\frac{5}{3}\lambda_0$. This then gives 
\begin{equation}
K_{a\gamma\gamma}= N_g(x +1/x)[\frac{m_u}{m_u+m_d}].
\end{equation}

\section{Invisible Axion Models}

The original PQ model \cite{PQ}, where $f_a=v_F$, was long ago ruled out by experiment. For example, one can estimate the branching ratio \cite{BPY}
\begin{equation}
BR(K^+ \to \pi^+ +a) \simeq 3 \times 10 ^{-5 }\lambda_0^2
=3 \times 10 ^{-5 } (x+1/x)^2
\end{equation} 
which is well above the KEK bound \cite{KEK}
$ BR(K^+\to \pi^+ + {\rm{ nothing}}) <3.8 \times 10 ^{-8}$. 
However, invisible axion models, where $f_a>>v_F$, are still viable.

Invisible axion models introduce scalar fields which carry PQ charge but are $SU(2)\times U(1)$ singlets. As a result, it is possible that the VEVs of these fields have a scale much larger than the one set by the weak interactions. Basically, two types of models have been proposed. The first of these models, due to
Kim \cite{Kim} and Shifman, Vainshtein and Zakharov \cite{SVZ} (the, so-called, KSVZ Model), introduces a scalar field $\sigma$ with $f_a= <\sigma> ~>> v_F$ and a superheavy quark $Q$ with $M_Q\sim f_a$ as the only fields carrying PQ charge. The second of these models, due to
 Dine, Fischler and Srednicki \cite{DFS} and Zhitnisky \cite{Z} (the, so-called, DFSZ Model), adds to the original PQ model a scalar field $\phi$ which carries PQ charge and $ f_a= <\phi> ~>> v_F$.

It is straightforward to repeat the calculations we just did above for these models, to get the axion mass and couplings. I
will do this explicitly here for the KSVZ model, because it is simple and illustrates well what we just did, and  will just quote the results for the DFSZ model. By construction,
the KSVZ axion does not interact with leptons and it only interacts with light quarks as the result of the strong and electromagnetic anomalies
\begin{equation}
L_{\rm{axion}}^{KSVZ}= \frac{a}{f_a}[\frac{g^2}{32\pi^2} F_a^{\mu\nu}\tilde{F}_{a\mu\nu} + 3e_Q^2\frac{\alpha}{4\pi}F^{\mu\nu}\tilde{F}_{\mu\nu}],
\end{equation}
where $e_Q$ is the em charge of the superheavy quark $Q$.

Since in the KSVZ model the ordinary Higgs do not carry PQ charge, the only interactions of the axion with the light quark sector come from the effective anomaly mass term, which here is given by
\begin{equation}
 L_{\rm{anomaly}}=-\frac{ (m^o_{\eta})^2}{2} [\eta + \frac{f_{\pi}}{2f_a} a ]^2\end{equation}
To the above one must add the standard quadratic term coming from the light quarks
\begin{equation}
L^{(2)}_{\rm{mass}}=-\frac{(m^o_{\pi})^2}{2}[\frac{m_u}{m_u+m_d}(\pi^o+\eta)^2+ \frac{m_d}{m_u+m_d}(\eta-\pi^o)^2].
\end{equation}
Diagonalizing $L_{\rm{anomaly}}$ and $ L^{(2)}_{\rm{mass}}$ gives
\begin{equation}
m_a= \frac{v_F}{f_a} \bar{m_a} \equiv \lambda_m \frac{v_F}{f_a} \bar{m_a}~;~\xi_{a\pi}=\lambda_3 \frac{f_{\pi}}{f_a}~;\xi_{a\eta}=\lambda_0\frac{f_{\pi}}{f_a},
\end{equation}
where for the KSVZ model these parameters are:
\begin{equation}
\lambda_m=1~ ;~\lambda_3=-\frac{m_d-m_u}{2 (m_u+m_d)}~ ;~ \lambda_0=-\frac{1}{2}.
\end{equation}
Note that, numerically, in the KSVZ model the axion mass is given by the formula:
\begin{equation}
m_a= \frac{v_F}{f_a} \bar{m_a} \simeq 6.3 [\frac{10^6 {\rm{GeV }}}{ f_a}]~{\rm{ eV}}.
\end{equation}

The calculation of $K_{ a\gamma\gamma}$ in this model is equally straightforward. To the contribution of the superheavy quark in the electromagnetic anomaly [$3e_Q^2$], one must add that coming from the mixing of the axion with the $\pi^o$ and the $\eta$š [ $\lambda_3 + \frac{5}{3}\lambda_0 $ ]. This gives, finally,
\begin{equation}
K _{a\gamma\gamma}= 3e_Q^2 - \frac{4m_d +m_u}{3(m_u+m_d)}.
\end{equation}

I will not go through the analogous calculation for the DFSZ model, but just quote the results. For this model, it proves convenient to define
\begin{equation} 
 X_1=\frac{2v_2^2}{v_F^2}~ ,~ X_2= \frac{2v_1^2}{v_F^2},
\end{equation} 
where $v_F= \sqrt{v_1^2 + v_2^2}$ and $v_1$ and $v_2$ are the two Higgs VEVs. Furthermore, if one rescales $f_a \to \frac{f_a}{2N_g}$ the axion mass in the DFSZ model is given by the same equation as for the KSVZ model, Eq. (56), corresponding to $\lambda_m=1$. With this rescaling understood, one then finds:
\begin{equation}
\lambda_3=\frac{1}{2}[\frac{X_1-X_2}{2N_g}- \frac{m_d-m_u}{m_d+m_u}]~~;~~ \lambda_0= \frac{1-N_g}{
2N_g},
\end{equation} 
and
\begin{equation}
K_{a\gamma\gamma}=\frac{4}{3}-\frac{4m_d +m_u}{3(m_d+m_u)}.
\end{equation}

Although the KSVZ and DFSZ axions are very light, very weakly coupled and very long-lived, they are not totally invisible. I will not describe in detail how these "invisible" axions affect astrophysics and cosmology, as others in this meeting have done so. However, in an attempt to make this brief precis on the strong CP problem and axions self contained, I will add a few comments on the bounds one derives for invisible axions.

Astrophysics gives bounds on the axion mass since axion emission, through šCompton production [$\gamma e \to ae$] and the Primakoff process, causes energy loss in stars.\cite{stars} The energy loss is inversely proportional to $ f_a^2$, and hence proportional to $m_a^2$. Thus axions must be light enough, so as not to affect stellar evolution. This is not the only way in which one obtains an upper bound on $m_a$. For instance, another bound on $m_a$ come from SN1987a, since axion emission through the process $ NN \to NNa$ in the core collapse affects the neutrino spectrum. \cite{SN} Typical bounds \cite{Raffelt} obtained from astrophysics require axions to be lighter than
\begin{equation}
 m_a  \leq ~1-10^{-3} {\rm{eV}}.
\end{equation}

Remarkably, cosmology gives a lower bound on the axion mass (and an upper bound on $ f_a$ ). \cite{PWW, AS, DF} The physics connected with this bound is simple to understand. When the Universe goes through the šPQ phase transition at $T \sim f_a >> \Lambda_{QCD}$ the QCD anomaly is ineffective. As a result, early in the Universe $ <a_{\rm{phys}}> $ is arbitrary. Eventually, when the Universe cools down to temperatures of order $T \sim \Lambda_{QCD} $ the axion obtains a mass and $<a_{\rm{phys}}> \to 0$. The PQ mechanism through which this happens, however, is not an instantaneous process and $ <a_{\rm{phys}}>$ oscillates to its final value. These coherent $\vec{p_a}=0$ axion oscillations contribute to the Universe's energy density and axions act as cold dark matter. The energy density contained in axion oscillations is proportional to $f_a$ and thus bounds on the energy density of cold dark matter in the Universe provide an upper bound on $f_a$ (and a lower bound on $m_a$).

I quote below the result of a recent calculation by Fox, Pierce and Thomas \cite{FPT} of the contribution to the Universe's energy density due to axions:
\begin{equation}
\Omega_ah^2š =0.5[\frac{f_a/\xi}{10^{12} {\rm{GeV}}}]^\frac{7}{6}[\theta_i^2 +(\sigma_{\theta})^2]\gamma.
\end{equation}
Here $\xi$ is the coefficient of the PQ anomaly and, with the way we defined $f_a$, $\xi=1$ for both the KSVZ and DFSZ models. The angle $\theta_i$ is the misalignment value for $ <a_{\rm{phys}}>/f_a$ and $\sigma_{\theta}$ is its mean squared fluctuation. Finally, $\gamma$ is a possible dilution factor for the energy density produced by axion oscillations. \cite{FPT}

One can use the recent WMAP bound on cold dark matter \cite{WMAP} to bound
\begin{equation}
\Omega_ah^2 \leq 0.12
\end{equation}
Assuming no dilution ($\gamma=1$) and using for $\theta_i$ an average misalignment angle $ \theta_i ^2= \frac{\pi^2}{3}$ and neglecting its fluctuations, the WMAP data gives the following cosmological bound for the PQ scale:
\begin{equation}
 f_a < 3 \times 10^{11} {\rm{GeV}}~~~ {\rm{ or}} ~~~ m_a > 2.1 \times 10^{-5} {\rm{ eV}}.
\end{equation}

\section{Concluding Remarks}

My conclusions are rather simple and telegraphic. In my view, after more than 25 years, the preferred solution to the strong CP problem remains still the idea that the Standard Model has an additional $ U(1)_{PQ}$ symmetry. Such a solution, necessarily, predicts the existence of a concomitant axion. Although Fermi scale axions have been ruled out, invisible axions models are still viable and axion oscillations toward its minimun could account for the dark matter in the Universe. This is an exciting possibility. However, no totally compelling invisible axion models exist, and there are no strong arguments to believe that $f_a$ takes precisely the value needed for axions to be the dark matter in the Universe. Nevertheless, it is encouraging that experimentalists are actively searching for signals of invisible axions.

\section*{Acknowledgments}

I would like to thank Markus Kuster for organizing this very nice workshop. Part of the material presented here is derived from my lectures at the KOSEF-JSPS Winter School on Recent Developments in Particle and Nuclear Theory, Seoul, Korea (1996) which appeared in J. Korean Phys. Soc. (Proc. Suppl.) {\bf 29}, S199 (1996). This work was supported in part by the Department of Energy under Contract No. FG03-91ER40662, Task C.

\end{document}